\documentclass[a4paper,fleqn,longmktitle]{cas-sc}

\usepackage[numbers]{natbib}
\usepackage{soul}
\usepackage[framemethod=tikz]{mdframed}
\usepackage{sidecap}

\begin{document}
\let\WriteBookmarks\relax
\def\floatpagepagefraction{1}
\def\textpagefraction{.001}
\shorttitle{}
\shortauthors{H.-X. Jiang, C.-R. Cai, J.-Q. Zhang, and M. Tang}

\title [mode = title]{The effect of role-based resource allocation on epidemic dynamics}                      

\author[1]{Hao-Xiang Jiang}[style=chinese]
\credit{Conceptualization, Methodology, Software, Validation, Data curation, Writing - Original Draft}
\author[1,2]{Chao-Ran Cai\corref{cor}}[style=chinese,orcid=0000-0002-1627-1564]
\cormark[1]
\ead{ccr@nwu.edu.cn}
\cortext[cor]{Corresponding author}
\credit{Conceptualization, Methodology, Validation, Formal analysis, Resources, Data curation, Writing - Review \& Editing, Funding acquisition}
\author[3]{Ji-Qiang Zhang}[style=chinese]
\credit{Validation, Formal analysis, Writing - Review \& Editing}
\author[4]{Ming Tang}[style=chinese]
\credit{Formal analysis, Writing - Review \& Editing, Funding acquisition}
\address[1]{School of Physics, Northwest University, Xi'an 710127, China}
\address[2]{Shaanxi Key Laboratory for Theoretical Physics Frontiers, Xi'an 710127, China}
\address[3]{School of Physics, Ningxia University, Yinchuan 750021, China}
\address[4]{School of Physics, East China Normal University, Shanghai 200241, China}

\begin{abstract}
We propose a coupled dynamical model of resource allocation and epidemic spread, inspired by the hierarchical structure of real-world therapeutic resource allocation. In this framework, network nodes are assigned distinct roles as either resource allocators or resource recipients.
As the average number of links per recipient from allocators increases, the prevalence exhibits one of four distinct response patterns across conditions: monotonically increasing, monotonically decreasing, U-shaped trend, or a sudden decrease with large fluctuations.
A mechanistic analysis uncovers three central insights: (i) a trade-off between efficient resource allocation and infection risk faced by allocators, (ii) the critical need to avoid resource redundancy when therapeutic efficiency is high, and (iii) the emergence of cascade-induced bistability in the coupled system.
\end{abstract}


\begin{highlights}
\item Analyze disease spread through role-based resource allocation
\item Allocators face a trade-off between efficient resource allocation and infection risk
\item Under high therapeutic resource efficiency, it’s crucial to avoid resource redundancy
\item Feedback between resource allocation and disease spread triggers a cascade, causing bistability
\end{highlights}

\begin{keywords}
Resource allocation \sep Epidemic spreading \sep Microscopic Markov chain theory \sep Role division \sep Network
\end{keywords}

\maketitle

\section{Introduction}

In the context of addressing infectious disease outbreaks~\cite{ZHONG20031353,GIRARD20104895,ijerph17051729}, the construction of a reasonable and effective resource allocation model holds paramount significance~\cite{Worby2020,Rossman2021}.
Public health frameworks operationalize resource deployment across three domains:
Primary prevention via preventive resources (vaccines establishing herd immunity);
Secondary containment using protective resources (masks reducing aerosol exposure);
Tertiary treatment applying therapeutic resources (antivirals mitigating severe outcomes).
The specificity of resources leads to the formation of diverse dynamic models between resource allocation and epidemic spreading. For instance, therapeutic resources contribute to reducing the spread of diseases but face dilution effects during outbreaks~\cite{WANG20191,Negi2025}: growing demand lowers individual access, establishing self-reinforcing cycles that either escalate or contain epidemics.

Early studies generally operated under the assumption that resources are fixed, static, and entirely separate from the dynamic epidemic process.
This fixed resource allocation rely on centralized control and can be termed budget-based resource allocation.
In relation to therapeutic resources, many studies assume a positive correlation between recovery probabilities and allocated resource quantities~\cite{PhysRevE.96.012321}.
This then transforms the problem of resource allocation into a direct study of the optimal recovery probability function, including linear \cite{https://doi.org/10.1002/rsa.20315}, exponential~\cite{10.1063/1.5049550,JIANG2018414}, power-law~\cite{PhysRevE.86.036114}, and other nonlinear functions~\cite{PhysRevE.96.012321}.

Following the initial studies, subsequent research extended its scope to delve into the interaction between network topology and the spread process~\cite{SUN2022112734}.
Resource allocation occurs through network links and targets immediate neighbors, as such we classify it as neighbor-based resource allocation.
A strand of research proposes a Resourced-No-resourced-Resourced framework~\cite{SUN2022112734,https://doi.org/10.1155/2021/6629105}, analogous to the Susceptible-Infected-Susceptible model, and couples it with disease transmission dynamics to study the interaction between resource allocation and epidemic spread.
However, these studies do not have the issue of resource consumption. 
These resources are applicable to information resources, and the overall architecture of this model is basically the same as that of another related model~\cite{PhysRevLett.111.128701,PhysRevResearch.5.013196,PhysRevResearch.5.033220}.
By introducing resource consumption, another type of neighbor-based resource allocation studies the self-organizing behavior of resource allocation on complex networks~\cite{PhysRevResearch.3.013157,10.1063/5.0227075,Huo2025} or their metapopulation versions~\cite{PhysRevE.105.064308,ZHANG2025115672}. 
These studies have particularly focused on therapeutic resources and have identified some dynamic mechanisms that are relevant to real world situations. 
For example, as the number of newly infected individuals continues to increase, medical resources will be over-consumed in the treatment of these patients, ultimately leading to the collapse of the medical system~\cite{doi:10.1056/NEJMp2006141}.

In the research on neighbor-based resource allocation, a fundamental and recurring assumption is that susceptible individuals serve as the sources of resources. 
For example, it is assumed that each susceptible node generates one unit resource at each time step~\cite{Chen_2018,power_info_SEIS}.
From the standpoint of certain resources—such as financial assets~\cite{Bttcher2015}—this assumption is reasonably realistic. 
However, when applied to therapeutic resources such as medicines, it becomes impractical due to constraints in production, distribution, and availability.
One characteristic of therapeutic resource allocation is the strict distinction between the resource allocators (such as doctors and pharmacists) and the resource recipients (such as patients).

In this paper, we introduce the concept of role-based resource allocation, in which network nodes are functionally segregated into resource allocators and resource recipients, and investigate its impact on disease spread.
Our role-based resource allocation differs from budget-based or neighborhood-based schemes, as it captures the characteristic of therapeutic resource allocation.
We systematically investigated the impact of the proportion of resource allocators and their average degree on the spread of infectious diseases across the four combinations of baseline recovery probability and treatment efficiency (high/high, high/low, low/high, low/low).

\section{Model}\label{model}
We focus exclusively on therapeutic resources, as opposed to preventive and protective resources, for analytical tractability and scope definition.
To ensure accessibility, resources are centrally coordinated and channeled through pharmacies, hospitals, and clinics.
In light of this consideration, we categorize the nodes within the network into two distinct classes: resource recipients that merely consume resources (also called general nodes, designated as \~{G}) and resource allocators that engage in resource allocation (designated as \~{D}), as exemplified by doctors and pharmacists.
In the context of epidemic spreading, we utilize the classical susceptible-infected-susceptible (SIS) model, in which S denotes the state of being susceptible and I indicates the state of being infected.
Mathematically, there are four state combinations: infected recipient (\~{G}I), susceptible recipient (\~{G}S), susceptible allocator (\~{D}S), and infected allocator (\~{D}I).
Note that letters with wavy lines (\~{G} and \~{D}) denote mutually non-convertible types, whereas ordinary letters (S and I) represent individual states that can be converted among themselves.

In our work, the following three assumptions are employed.
Firstly, there are no \~{D}-\~{D} links since, in reality, the connections among medical institutions are scarce.
Secondly, resources can only be distributed through \~{D}S-\~{G}I links. The rationale lies in the fact that \~{G}S individuals have no resource demands, and \~{D}I individuals have suspended resource distribution to prevent cross-infection. Furthermore, this action also cuts off the transmission pathway of the disease for individual \~{D}I.
Thirdly, within each time step, each \~{D}-type individual will receive $\frac{1}{r}$ resources, where $r$ is the fraction of \~{D} nodes. 
The distribution of these resources occurs as follows: if \~{D}S-\~{G}I links exist, the resources are evenly allocated through them; otherwise, the resources remain undistributed.
Here, the total resources generated by the system at each step amount to $N$, which is equivalent to the network size.

The available resources to individuals of type \~{G}S, type \~{G}I, type \~{D}S, and type \~{D}I are defined as $R_i^{\mathrm{\tilde{G}S}}(t)$, $R_i^{\mathrm{\tilde{G}I}}(t)$, $R_i^{\mathrm{\tilde{D}S}}(t)$, and $R_i^{\mathrm{\tilde{D}I}}(t)$, respectively.
Overall, the resource allocation of various types of individuals is as follows
\begin{equation}
  \begin{aligned}\label{ri}
R_i^{\mathrm{\tilde{G}I}}(t)&=\sum_{j\in \Omega(i)\cap\Omega(\mathrm{\tilde{D}S})} \frac{1}{rm_j^{\mathrm{\tilde{G}I}}},\\ 
R_i^{\mathrm{\tilde{D}I}}(t)&=\frac{1}{r},\\ 
R_i^{\mathrm{\tilde{G}S}}(t)&=0,\\
R_i^{\mathrm{\tilde{D}S}}(t)&=
\begin{cases} 
0, & \text{if at least one neighbor is infected}, \\
\frac{1}{r}, & \text{if there are no infected neighbors}.
\end{cases}
\end{aligned}
  \end{equation}
Here, $\Omega(i)$ is defined as the set of neighbors of individual $i$, $\Omega(\mathrm{\tilde{D}S})$ denotes the set of all \~{D}S individuals, and $m_j^{\mathrm{\tilde{G}I}}$ represents the number of \~{G}I individuals among the neighbors of the \~{D}S individual $j$. 

\begin{figure}
  \centering
  \includegraphics[width=\linewidth]{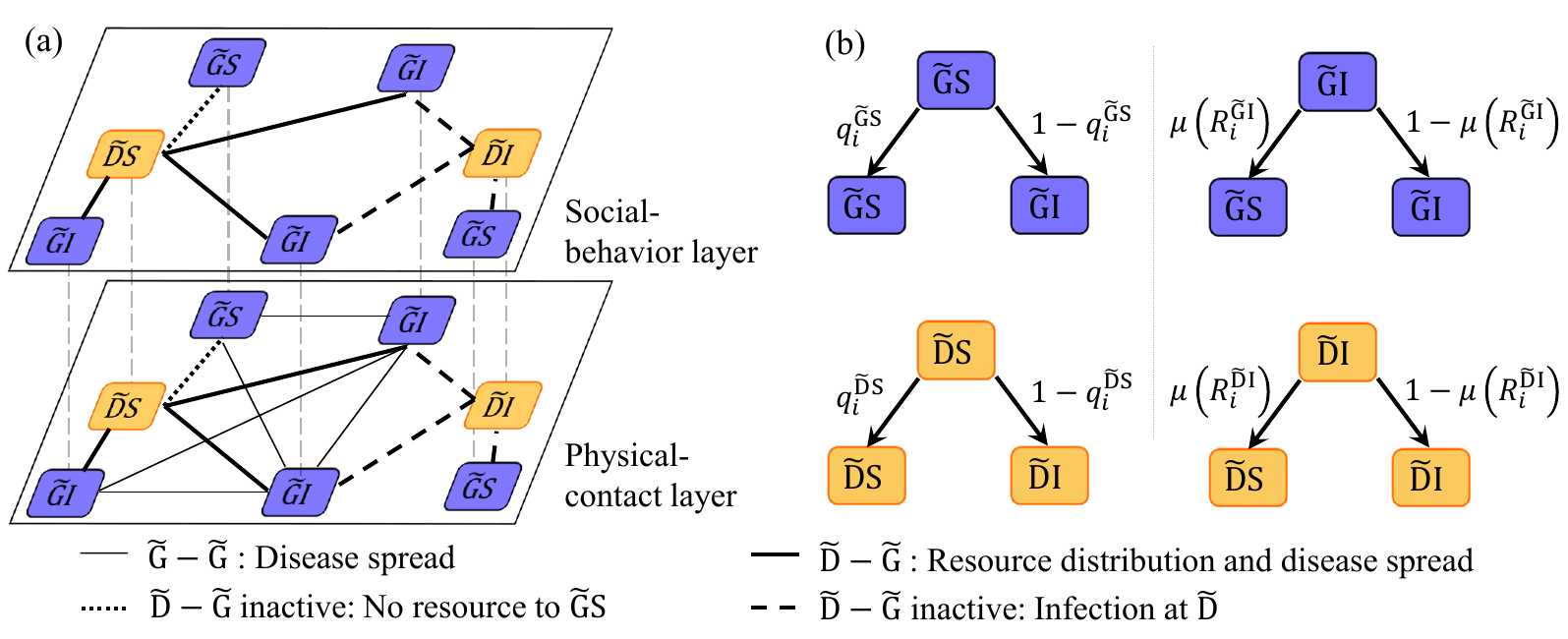}
  \caption{(a) Schematic illustration of the resource-epidemic model with role division. In the social-behavior layer, the thick solid lines are employed for resource distribution. In the physical-contact layer, the thick and thin solid lines are used for disease spread. (b) The transition probability trees are presented for the classes \~{G}I, \~{G}S, \~{D}I, and \~{D}S, respectively. 
 \label{netpic}}
\end{figure}

The corresponding steps for generating the static network are as follows:
\begin{enumerate}
  \item A number $rN$ of nodes is randomly selected to be classified as type \~{D}, while the remaining nodes are classified as type \~{G}. 
  \item  A total of $\frac{\langle k_1\rangle}{2}(1-r)N$ \~{G}-\~{G} links are randomly created among individuals of type \~{G} in the physical-contact layer. Here, $\langle k_1\rangle$ denotes the average number of links that a \~{G} node has with other \~{G} nodes.
  \item  A total of $\langle k_2\rangle(1-r)N$ identical \~{D}-\~{G} links are randomly created on both the physical-contact layer and social-behavior layer simultaneously. Here, $\langle k_2\rangle$ denotes the average number of links that a \~{G} node has with \~{D} nodes.
\end{enumerate}
Therefore, the average degrees of individual \~{G} and individual \~{D} are $\langle k_1\rangle+\langle k_2\rangle$ and $\langle k_2\rangle(1-r)/r$, respectively.
The physical-contact layer represents the disease sprading network, whereas the social-behavior layer is a subnetwork of it that serves as the resource allocation network.
The schematic illustration of the network is presented in Fig.~\ref{netpic}(a).

For epidemic spreading, an individual in the susceptible state, whether \~{G}S or \~{D}S, becomes infected with probability $\beta$ upon contact with a neighboring infected individual of type \~{G}I. 
Meanwhile, an infected individual, either \~{G}I or \~{D}I, recovers with probability of $\mu(R_i)$.
In formulating the relation between the recovery probability and resources, three practical scenarios need to be taken into account. 
Firstly, even when resources are absent, individuals may retain a certain probability of recovery due to their inherent immunity. 
Secondly, there is a positive correlation between the recovery probability and the allocated resource quantities~\cite{PhysRevE.96.012321}.
Thirdly, there exists a saturation effect of resources on enhancing the recovery probability~\cite{PhysRevResearch.3.013157,power_info_SEIS,PhysRevE.111.044301}. 
Therefore, this relation is primarily characterized by three physical quantities: the baseline recovery probability (under zero-resource conditions), the saturated recovery probability (under infinite-resource conditions), and the treatment efficiency (which measures the rate of approaching the saturation state).
Given these considerations, we propose the following relation:
\begin{equation}
  \mu(R_i)=\frac{\mu_0}{1+\frac{1}{\alpha}\exp(-\omega R_i)},\label{mu}
\end{equation}
where $\alpha>0$ and $\omega>0$ can serve as a metric for the strength of the baseline recovery probability and the treatment efficiency, respectively.
Equation~\eqref{mu} shows that the baseline recovery probability and the saturated recovery probability are respectively $\frac{\alpha}{1+\alpha}\mu_0$ and $\mu_0$.
The sigmoid function defined in Eq.~\eqref{mu} is plotted in Fig.~\ref{figapp}(a) of the~\ref{supa}.
Equation~\eqref{mu} is selected for this paper, although it is not the only one that satisfies the requirements; alternative formulations, such as those presented in Refs.~\cite{PhysRevResearch.3.013157,PhysRevE.105.064308,CHEN2019156}, are also capable of generating similar relation curves.

\section{Theoretical analysis }\label{Theory}
We employ the microscopic Markov chain theory (widely used to analyze disease dynamics on networks~\cite{10.1063/5.0125873,10.1063/5.0151881,GAO2026117588,ZHANG2026117775}) to conduct an analysis of the dynamics of epidemic spread in the context of resource allocation.
Let $\textbf{G}=[g_{ij}]$ denote the adjacency matrix of a sub-network that only contains \~{G}-\~{G} links. Specifically, for any two individuals $i$ and $j$, if there exists a \~{G}-\~{G} link between them, then $g_{ij}=1$; conversely, if there is no such link, then $g_{ij}=0$.
Consider $\textbf{D}=[d_{ij}]$ as the adjacency matrix corresponding to a sub-network that only contains \~{D}-\~{G} links.

Define $\rho_i^{\mathrm{\tilde{G}S}}(t)$, $\rho_i^{\mathrm{\tilde{G}I}}(t)$, $\rho_i^{\mathrm{\tilde{D}S}}(t)$, and $\rho_i^{\mathrm{\tilde{D}I}}(t)$ as the probabilities that individual $i$ occupies the states of \~{G}S, \~{G}I, \~{D}S, and \~{D}I, respectively, at time step $t$.
These probabilities satisfy the following condition,
\begin{equation}\label{e3}
 \left(\rho_i^{\mathrm{\tilde{G}S}}+\rho_i^{\mathrm{\tilde{G}I}}=0\land\rho_i^{\mathrm{\tilde{D}S}}+\rho_i^{\mathrm{\tilde{D}I}}=1\right)\oplus \left(\rho_i^{\mathrm{\tilde{G}S}}+\rho_i^{\mathrm{\tilde{G}I}}=1\land\rho_i^{\mathrm{\tilde{D}S}}+\rho_i^{\mathrm{\tilde{D}I}}=0\right)\Leftrightarrow \mathrm{True},
\end{equation}
where $\oplus$ is the XOR operator in logic. Equation~\eqref{e3} indicates that there are only two independent variables.

In accordance with Eqs.~\eqref{ri}-\eqref{mu}, we can deduce that when individual $i$ occupies the state \~{D}I, its recovery probability is given by 
\begin{equation}
\mu(R_i^{\mathrm{\tilde{D}I}})=\frac{\mu_0}{1+\frac{1}{\alpha}\exp(-\frac{\omega}{r})}.
\end{equation}
As individual $j$ is in the \~{G}I state, its recovery probability is related to the \~{D}S neighbors and the \~{G}I individuals that are directly connected to these neighbors.
The probability distribution of individual $j$ obtaining the resource $\frac{1}{rm_{j'}}$ from a given \~{D}S-state neighbor $j'$ is 
\begin{equation}\label{p}
p\left(X_{j'}=\frac{1}{rm_{j'}}\right) = \sum_{\substack{H\subseteq\Omega(j')\setminus \{j\} \\ \vert H\vert=m_{j'} - 1}}\prod_{i\in H}\rho_i^{\mathrm{\tilde{G}I}}\prod_{i\in (\Omega(j')\setminus \{j\})\setminus H}(1 - \rho_i^{\mathrm{\tilde{G}I}}).
\end{equation}
Here, $\vert H\vert$ represents the number of elements in the set $H$, $m_{j'}\in[1,k_{j'}]$ is the number of \~{G}I neighbors of individual $j'$, and $k_{j'}$ is the degree of individual $j'$.
Subsequently, the probability distribution of individual $j$ acquiring the complete set of resources from its neighbors can be represented as
\begin{equation}\label{P}
P(Y)=\sum_{\substack{F\subseteq\Omega(j)}}f_{F}\cdot g_{F}(Y),
\end{equation}
where
\begin{equation}
f_{F}= \prod_{i\in F}(1 - \rho_i^{\mathrm{\tilde{D}I}})\prod_{i\in \Omega(j)\setminus F}\rho_i^{\mathrm{\tilde{D}I}} \qquad\text{and}\qquad
g_{F}(Y)=P\left(Y=\sum_{j'\in F}X_{j'}\right).
\end{equation}
Here, $f_{F}$ denotes the probability of the occurrence of the set $F$. 
Meanwhile, $g_{F}(Y)$ represents the conditional probability of the total resource $Y$ given that the set $F$ occurs. 
Additionally, $X_{j'}$ stands for the resource provided by neighbor $j'$, and its distribution is shown in Eq.~\eqref{p}.
Finally, the corresponding average recovery probability of individual $j$ is 
\begin{equation}\label{e8}
\langle\mu(R_j^{\mathrm{\tilde{G}I}})\rangle=\sum_Y\frac{\mu_0P(Y)}{1+\frac{1}{\alpha}\exp(-\omega Y)}.
\end{equation}
It should be noted that the theories presented in Refs.~\cite{PhysRevResearch.3.013157,Chen_2018,power_info_SEIS} do not guarantee equality between the expected total amount of allocated resources and that of received resources. 
In contrast, our Eqs.~\eqref{p}-\eqref{e8} ensure this balance during the calculation of the average recovery probability, thereby overcoming the limitations of prior approaches.

When individual $i$ and individual $j$ are in the \~{D}S and \~{G}S states, the infected probabilities denote as $q_i^{\mathrm{\tilde{D}S}}$ and $q_j^{\mathrm{\tilde{G}S}}$ respectively, are given by the following equations:
\begin{align}
q_i^{\mathrm{\tilde{D}S}}&=1-\prod_k(1-d_{ik}\rho_k^{\mathrm{\tilde{G}I}}\beta),\label{qd}\\
q_j^{\mathrm{\tilde{G}S}}&=1-\prod_k(1-g_{jk}\rho_k^{\mathrm{\tilde{G}I}}\beta).\label{qs}
\end{align}
Figure~\ref{netpic}(b) illustrates the transition probabilities of individual $i$ across possible states at each time step.
Then, the dynamic equations for each individual can be written as
\begin{align}
\rho_i^{\mathrm{\tilde{D}I}}(t+1)=&q_i^{\mathrm{\tilde{D}S}}[1-\rho_i^{\mathrm{\tilde{D}I}}(t)]+[1-\mu(R_i^{\mathrm{\tilde{D}I}})]\rho_i^{\mathrm{\tilde{D}I}}(t),\label{tD}\\
\rho_j^{\mathrm{\tilde{G}I}}(t+1)=&q_j^{\mathrm{\tilde{G}S}}[1-\rho_j^{\mathrm{\tilde{G}I}}(t)]+[1-\langle\mu(R_j^{\mathrm{\tilde{G}I}})\rangle]\rho_j^{\mathrm{\tilde{G}I}}(t),\label{tA}
\end{align}
where $i\neq j$, given that individuals classified as type \~{G} and those categorized as type \~{D} inherently represent distinct entities.
The fraction of infected individuals at time $t$, denoted as $\rho(t)$, can be calculated from
\begin{equation}
\rho(t)=\frac{1}{N}\left[\sum_i\rho_i^{\mathrm{\tilde{D}I}}(t)+\sum_j\rho_j^{\mathrm{\tilde{G}I}}(t)\right].
\end{equation}

Near the critical transmission probability $\beta_c$, we set $\rho_j^{\mathrm{\tilde{G}I}}=\epsilon_j\ll1$.
Note that the consideration of $\rho_i^{\mathrm{\tilde{D}I}}$ is unnecessary since individuals of type \~{D}I is incapable of disease spread.
Neglecting high-order terms of $\epsilon_j$, Eq.~\eqref{qs} is rewritten as
\begin{equation}\label{liner} 
    q_j^{\mathrm{\tilde{G}S}}=\beta\sum_kg_{jk}\epsilon_k.
\end{equation}
Substituting Eq.~\eqref{liner} into Eq.~\eqref{tA}, we obtain
\begin{equation}\label{Aliner}
\sum_k \left[\beta g_{jk}-\delta_{jk}\langle\mu(R_j^{\mathrm{\tilde{G}I}})\rangle\right]\epsilon_k=0,
\end{equation}
where $\delta_{jk}$ is the element of the identity matrix.
Here, $\langle\mu(R_j^{\mathrm{\tilde{G}I}})\rangle$ is related to $\langle k_2\rangle$. While a general threshold expression cannot be provided, explicit expressions can be derived for the two limiting cases.

In cases where $\langle k_2\rangle$ is relatively large, resources are nearly evenly distributed, leading to the relation $R_j^{\mathrm{\tilde{G}I}}=1/\epsilon_j\to \infty$. 
It indicates that resources are highly abundant and $\langle\mu(R_j^{\mathrm{\tilde{G}I}})\rangle\approx \mu_0$.
Then, one can obtain the epidemic threshold
\begin{equation}\label{betac0}
    \beta_c=\frac{\mu_0}{\Lambda_{\rm}(\textbf{G})},
\end{equation}
where $\Lambda_{\rm}(\textbf{G})$ is the largest eigenvalue of the matrix $\textbf{G}$.

When $\langle k_2\rangle=0$ or is sufficiently small, \~{D}-\~{G} links are absent or sparse, leading to the opposite extreme where $\langle\mu(R_j^{\mathrm{\tilde{G}I}})\rangle\approx \frac{\alpha}{1+\alpha}\mu_0$. In this case, the threshold is
\begin{equation}\label{betac1}
    \beta_c=\frac{\alpha}{1+\alpha}\frac{\mu_0}{\Lambda_{\rm}(\textbf{G})}.
\end{equation}

In fact, in these two extreme cases, the model degenerates into the classic SIS model, and Eqs.~\eqref{betac0} and \eqref{betac1} can also be derived via first-order approximation using the spectral approach~\cite{PhysRevLett.109.128702,w91q-hm26}.

\section{Results and discussion}\label{result}
\begin{figure*}
	\includegraphics[width=\linewidth]{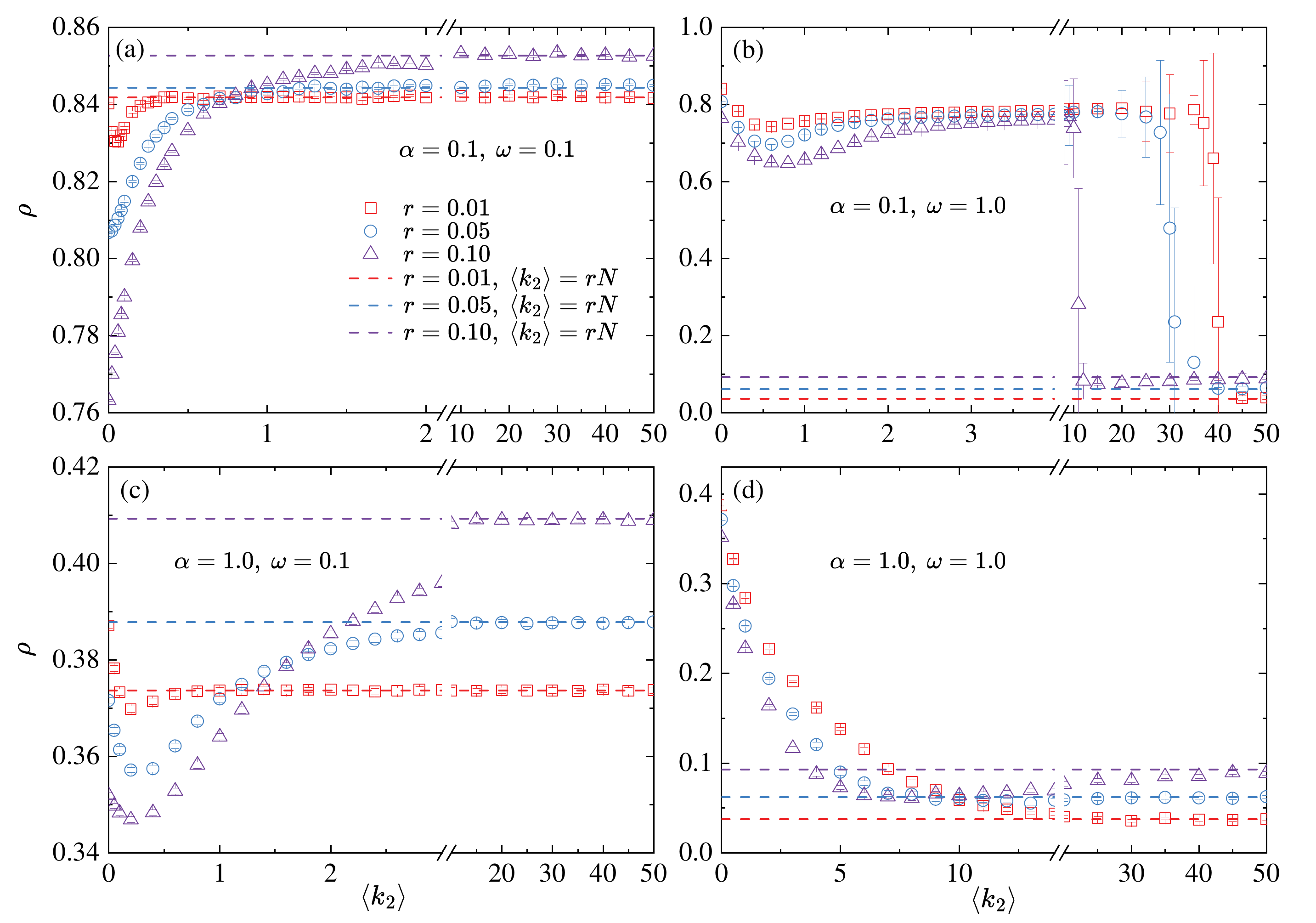}
	\caption{The impact of the fraction of resource allocators ($r$) and the average number of connections from recipients to allocators ($\langle k_2\rangle$) on the prevalence ($\rho$) in the inevitable scenario of a disease outbreak. Parameters: $\beta=0.12$, $\mu_0=0.5$, initial fraction of infected individuals $\rho(0)=0.01$.}\label{fig2}
\end{figure*}

\subsection{Instructions for parameter selection}
Unless explicitly stated, we fix from now on the total population size $N=10^5$, average degree of the epidemic layer $\langle k_1\rangle=4$, and $\mu_0=0.5$.
Each simulation was run for 1000 iterations. The long run equilibrium results (shown in Fig.~\ref{fig2}, Fig.~\ref{fig3}, and Fig.~\ref{fig6}) represent the average of frequencies over the last 100 iterations in 100 independent simulations.

In practical scenarios, individuals generally establish connections with only a limited number of medical institutions.
However, as a model study, we can arbitrarily change the value of $\langle k_2\rangle$ to fully capture the essence of the phenomenon in the research.
In this paper, the variable $\langle k_2\rangle$ is assigned a range of $\langle k_2\rangle\in [0,50]$, with the limiting case where $\langle k_2\rangle=rN$ also being considered for reference.
In this limiting case, each individual of type \~{G} has connections with all individuals of type \~{D}.

For the fraction of individual \~{D}, we choose three values $r=0.01, 0.05$ and $0.1$.

In terms of infection probability $\beta$, three distinct scenarios can be discerned.
Firstly, even in the case of unlimited resource availability, the disease can still break out.
This situation is called general infection region and corresponds to Sec.~\ref{reb} (setting $\beta=0.12$). Secondly, the disease fails to spread when resources are abundant but can disseminate when resources are entirely depleted.
This scenario is called weak infection region and corresponds to Sec.~\ref{rec} (with $\beta=0.09$ employed). Lastly, in the scenario where resources are completely absent, the disease remains unable to spread.

For the strength of the baseline recovery probability and the treatment efficiency, we consider two values for each parameter: $0.1$ and $1$. Specifically, $\alpha=0.1$ signifies a weak baseline recovery probability, whereas $\alpha=1$ denotes a strong one. Similarly, $\omega=0.1$ implies a low treatment efficiency, and $\omega=1$ indicates a high treatment efficiency.

See the~\ref{supa} for more details on the selection of parameters $\alpha$, $\omega$, and $\beta$.

\subsection{General infection region}\label{reb}

In this subsection, the parameters we have specified are such that they allow for the possibility of a disease outbreak even under conditions of infinite resource availability.
In Fig.~\ref{fig2}, the prevalence $\rho$ is plotted as a function of the average number of allocators contacted by recipients, with separate curves showing its behavior for various the fraction of resource allocators.
Overall, as the parameter $\langle k_2\rangle$ increases, the prevalence $\rho$ will gradually converge to the value corresponding to the fully-connected state of the resource layer, which is in line with expectations.
As the parameter $r$ increases, the prevalence $\rho$ exhibits a decreasing trend when the parameter $\langle k_2\rangle$ is small, whereas it shows an increasing trend when $\langle k_2\rangle$ is large.
Moreover, Fig.~\ref{fig2} illustrates four distinct behaviors of $\rho$ as the parameter $\langle k_2\rangle$ increases: {\it monotonically increasing}, {\it monotonically decreasing}, {\it initially decreasing and then increasing}, and {\it a sudden decrease with large fluctuations}.
Meanwhile, as $r$ increases, Fig.~\ref{fig2}(a) demonstrates that the curve’s behavior transitions from the pattern of initially decreasing and then increasing to the pattern of monotonically increasing. 
Similarly, Fig.~\ref{fig2}(d) shows that the curve’s behavior changes from the pattern of monotonically decreasing to the pattern of initially decreasing and then increasing.

\begin{figure}
	\includegraphics[width=\linewidth]{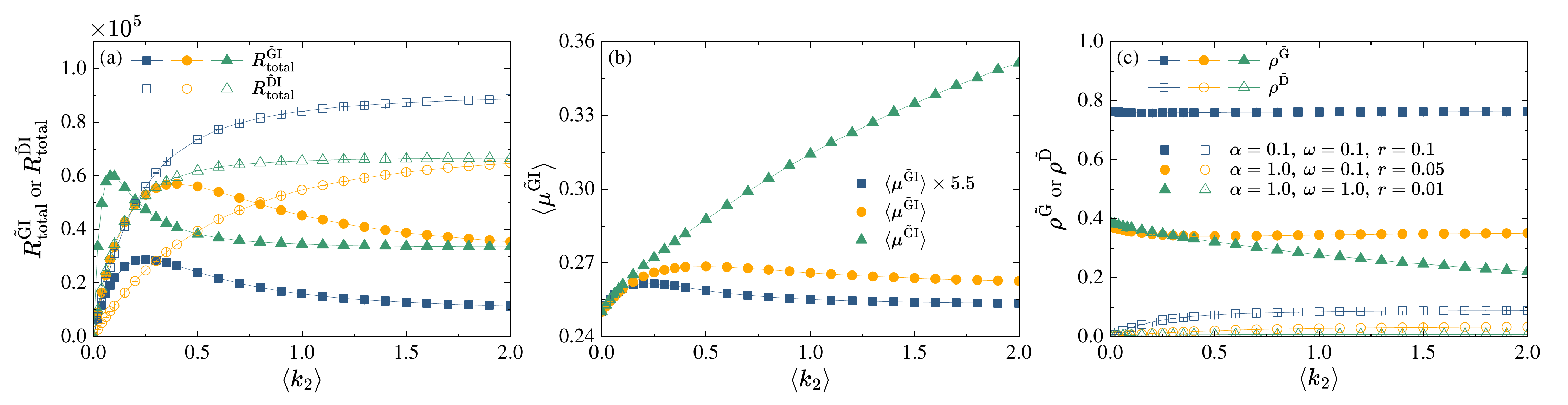}
	\caption{The total resources in (a), average recovery probability in (b), and prevalence in (c) across different categories as functions of the variable $\langle k_2\rangle$. The parameters employed in square, circle, and triangle scatters correspond respectively to the monotonically increasing, initially decreasing and subsequently increasing, and monotonically decreasing trends observed in Fig.~\ref{fig2}.}\label{fig3}
\end{figure}

{\bf Dynamical mechanism I:} When no connections exist between \~{D} and \~{G} (i.e., $\langle k_2\rangle = 0$), all the resources are concentrated among \~{D}S individuals. 
As $\langle k_2\rangle$ increases, \~{D}S individuals allocate resources to \~{G}I neighbors through \~{D}S-\~{G}I links. 
Meanwhile, the risk of these \~{D}S individuals getting infected rises.
This rise in \~{D}I individuals subsequently results in a decrease in the resources allocated to \~{G}I individuals. 
Physical mechanism I implies a trade-off between the efficient resources allocation and the infection risks faced by those resource allocators.

As shown in Fig.~\ref{fig3}(a), the total resources of \~{G}I individuals ($R_{\mathrm{total}}^{\mathrm{\tilde{G}I}}$) first increase and then decrease with increasing $\langle k_2\rangle$, while those of \~{D}I individuals ($R_{\mathrm{total}}^{\mathrm{\tilde{D}I}}$) exhibit a monotonic increase.
$R_{\mathrm{total}}^{\mathrm{\tilde{D}I}}$ is positively correlated with $\rho^{\mathrm{\tilde{D}}}$, the fraction of \~{D}I individuals, as shown by the hollow scatters in Fig.~\ref{fig3}, and their relation given by $R_{\mathrm{total}}^{\mathrm{\tilde{D}I}}=\rho^{\mathrm{\tilde{D}}}N/r$.
As predicted by Eq.~\eqref{mu}, $R_{\mathrm{total}}^{\mathrm{\tilde{G}I}}$ is positively correlated with $\langle\mu^{\mathrm{\tilde{G}I}}\rangle$, the average recovery probability of \~{G}I individuals, but negatively correlated with $\rho^{\mathrm{\tilde{G}}}$, the fraction of \~{G}I individuals. 
This general correlation pattern is manifested in the square and circular scatters in Fig.~\ref{fig3}. 
The above analysis explains the difference between the {\it monotonically increasing} pattern and the {\it initially decreasing and then increasing} in Fig.~\ref{fig2}, which depends on the relative magnitudes of the absolute slopes of $\rho^{\mathrm{\tilde{D}}}$ and $\rho^{\mathrm{\tilde{G}}}$ at small $\langle k_2\rangle$.
Therefore, a monotonically increasing pattern often occurs when $r$ is large and $\alpha$, $\omega$ are small, as large $r$ gives $\rho^{\mathrm{\tilde{D}}}$ a large absolute slope and small $\alpha$, $\omega$ give $\rho^{\mathrm{\tilde{G}}}$ a small one—this behavior is exemplified by the square scatters in Fig.~\ref{fig3}(c).

However, the triangle scatters in Fig.~\ref{fig3} exhibits a deviation from the expected behavior. 
This anomaly suggests the existence of an additional mechanism that leads to a decrease in the prevalence rate as the parameter $\langle k_2\rangle$ increases.
Notably, the unexpected downward trends occurred only in the case where $\omega=1.0$.
\begin{figure*}
	\includegraphics[width=\linewidth]{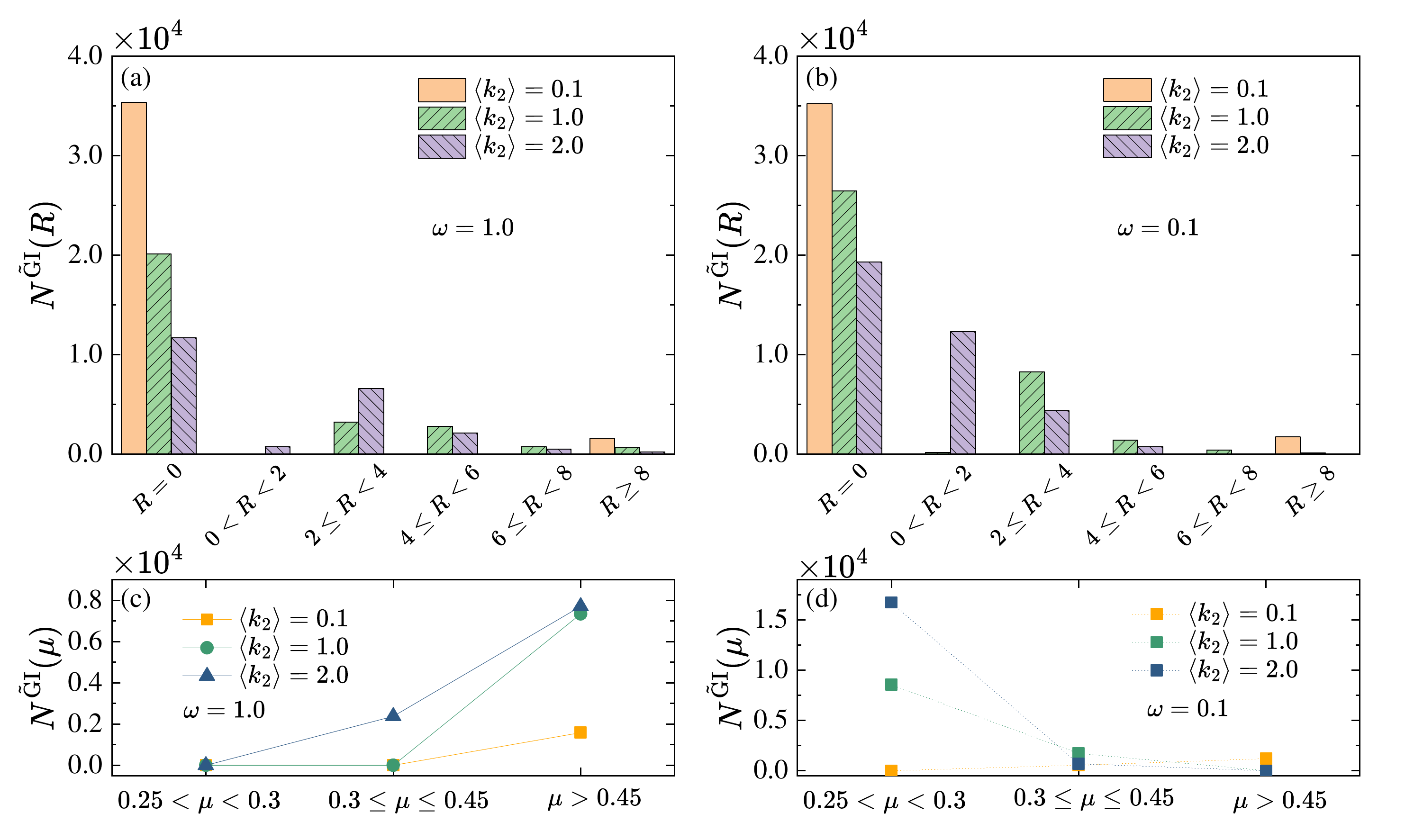}
	\caption{The number of \~{G}I individuals $N^{\mathrm{\tilde{G}I}}$ as functions of $R$ and $\mu$, respectively. Data are from a single simulation at the final time step ($t=1000$). Parameters: $\alpha=1.0$, $r=0.01$, $\beta=0.12$, $\mu_0=0.5$, and $\rho(0)=0.01$.}\label{fig4}
\end{figure*}

{\bf Dynamical mechanism II:} 
When links between \~{D} and \~{G} are sparse (small $\langle k_2\rangle$), a limited number of \~{G}I individuals capture a disproportionate resources, leaving the majority excluded due to limited allocation channels. As connectivity increases, resource distribution becomes more uniform: more \~{G}I individuals receive resources, but each receives less. 
Thus, physical mechanism II facilitates broader access through the reduction of allocative redundancy.
While more \~{G}I individuals gain resources—increasing their recovery pribabilities—others receive less than before, potentially reducing their recovery pribabilities. 
The net effect on average recovery probability depends on the competition between these two opposing trends.
\begin{figure}
	\includegraphics[width=\linewidth]{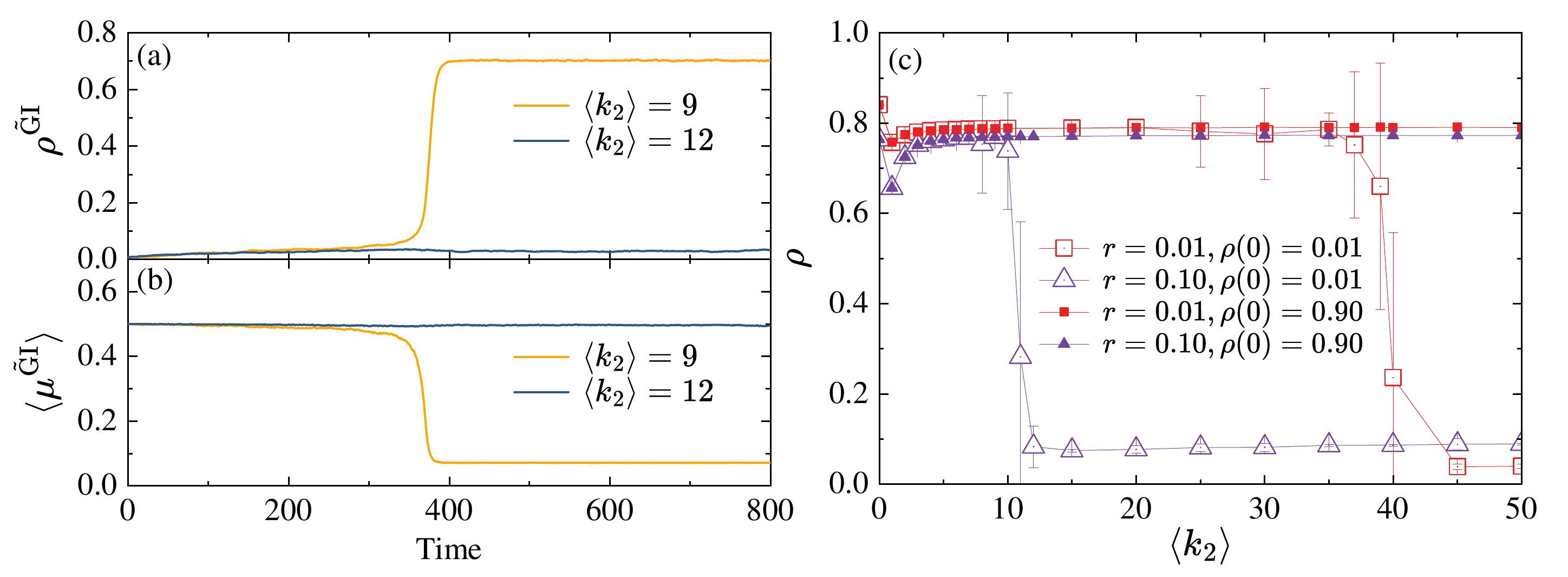}
	\caption{Time series for the fraction in (a) and the average recovery pribability in (b) of \~{G}I individuals. (c) The prevalence $\rho$ as a function of $\langle k_2\rangle$ with different initial infected individuals. Parameters: $\beta=0.12$, $\mu_0=0.5$, $\alpha=0.1$, $\omega=1.0$; (a) and (b) $r=0.1$, $\rho(0)=0.01$.}\label{fig5}
\end{figure}

As shown in Fig.~\ref{fig4}(a) and Fig.~\ref{fig4}(b), for $\langle k_2\rangle=0.1$, resource allocation among \~{G}I individuals is highly skewed, with most resources concentrated in two extreme regimes: individuals receiving no resources ($R=0$) or substantial allocations ($R\geq8$). 
This distribution indicates significant allocative redundancy. 
As $\langle k_2\rangle$ increases, the distribution of resources becomes increasingly uniform, with a marked decrease in the number of individuals receiving no resources. 
When compared with the scenario where $\omega=0.1$, the case of $\omega=1.0$ necessitates only a small quantity of resources to reach the saturation value of the recovery probability, see the Fig.~\ref{figapp}(a) of the~\ref{supa}. 
As a result, Fig.~\ref{fig4}(c) shows that for $\omega=1.0$, the number of \~{G}I individuals with recovery probabilities close to saturation [$\mu(R\to\infty)=0.5$] increases substantially with $\langle k_2\rangle$.
In contrast, Fig.~\ref{fig4}(d) shows that for $\omega=0.1$, those near the saturated probability gradually decrease to zero, while individuals near the baseline recovery probability [$\mu(R=0)=0.25$ for $\alpha=1.0$] rises markedly.
The above analysis indicates that physical mechanism II effectively suppresses disease spread, but only when treatment efficiency is sufficiently high (i.e., high $\omega$). 
This explains the {\it monotonically decreasing} pattern in Fig.~\ref{fig2} and forms the foundation for interpreting {\it a sudden decrease with large fluctuations}.

In Fig.~\ref{fig5}(a) and Fig.~\ref{fig5}(b), we have presented the time series analysis of the fraction of \~{G}I individual, denoted as $\rho^{\mathrm{\tilde{G}I}}$, along with the average recovery probability of those \~{G}I individuals, denoted as $\langle\mu^{\mathrm{\tilde{G}I}}\rangle$.
As is evident from Fig.~\ref{fig5}(b), initially, the average probabilities of the two lines were identical (both reaching the maximum value).
Subsequently, a slight disparity emerged and gradually widened.
For the lines corresponding to $\langle k_2\rangle=9$, at approximately time step $380$ (this value may vary for different runs), Fig.~\ref{fig5}(a) and Fig.~\ref{fig5}(b) show a sudden increase and a sudden drop, respectively.
Combining Fig.~\ref{fig5}(a) and Fig.~\ref{fig5}(b) reveals a negative correlation between overall recovery probability and the number of infected individuals, consistent with Ref.~\cite{PhysRevE.100.032310}.

{\bf Dynamical mechanism III:} 
Initially, the emergence of a small number of additional \~{G}I individuals possibly resulted from the imbalance in resource allocation.
This situation resulted in a decrease in the average resource and the average recovery probability of \~{G}I individuals.
Subsequently, this reduction further spurred an increase in the number of infected individuals.
This cycle repeated itself, ultimately giving rise to a cascading effect.
This explains the {\it a sudden decrease with large fluctuations} pattern in Fig.~\ref{fig2}.
Moreover, the above analysis indicates the emergence of a bistable region: depending on the initial seeds, the system evolves toward one of two stable steady states. Large initial seeds can trigger the cascade immediately, bypassing the slow buildup phase and leading directly to a high-prevalence steady state. In contrast, small initial seeds fail to initiate such a cascade and remain trapped in a low-prevalence equilibrium.
In Fig.~\ref{fig5}(c), we plot the prevalence $\rho$ as a function of $\langle k_2\rangle$ with different initial infected individuals.  A bistable behavior is observed after the region of {\it a sudden decrease with large fluctuations}.

\subsection{Weak infection region}\label{rec}
\begin{figure*}
	\includegraphics[width=\linewidth]{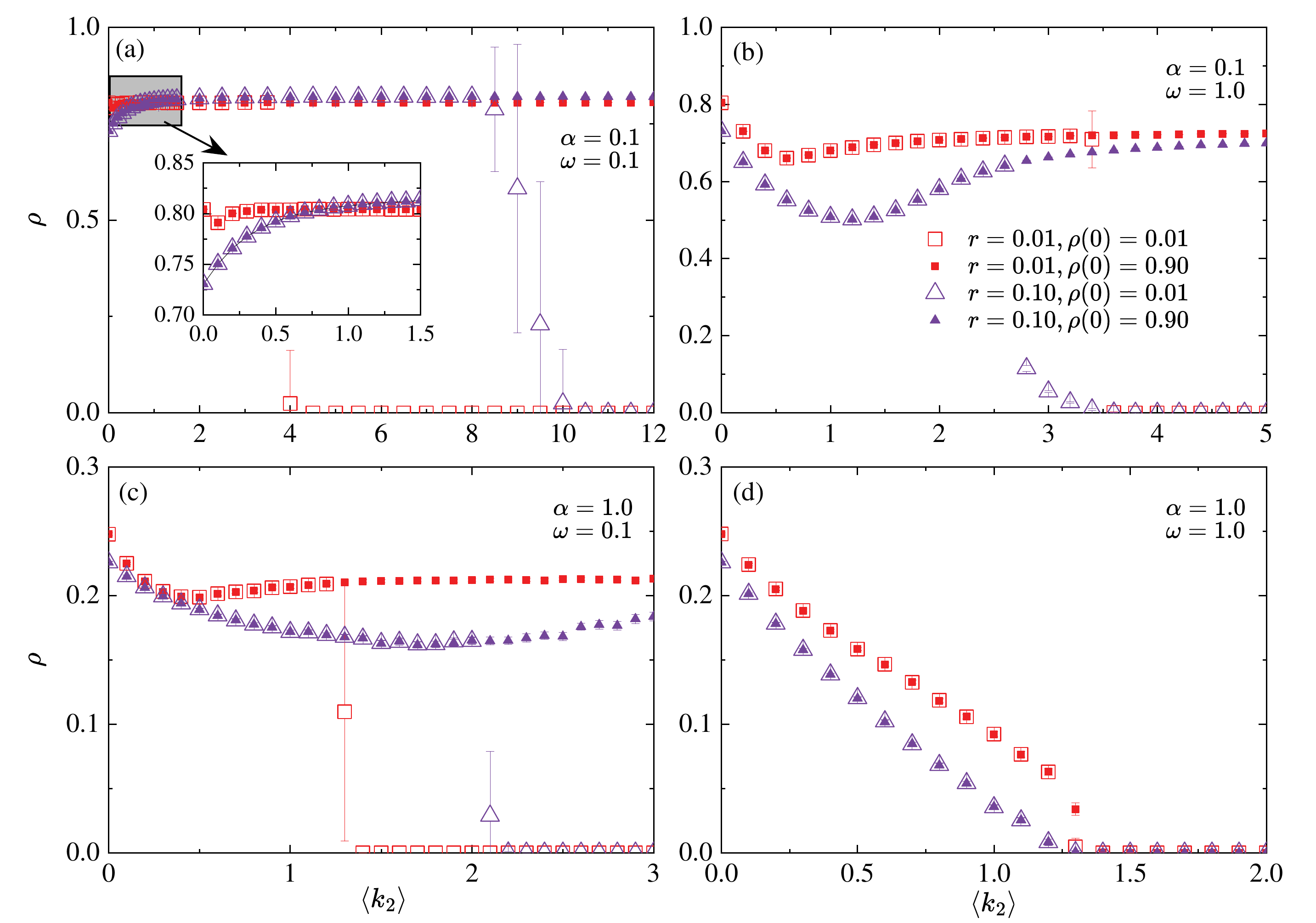}
	\caption{The impact of the fraction of resource allocators ($r$) and the average number of connections from recipients to allocators ($\langle k_2\rangle$) on the prevalence ($\rho$) under the condition that disease outbreaks may be prevented. Parameters: $\beta=0.09$, $\mu_0=0.5$.}\label{fig6}
\end{figure*}

In this subsection, the parameters we have set ensure that the disease cannot spread when resources are abundant; however, it can spread when resources are at zero.
In Fig.~\ref{fig6}, the prevalence $\rho$ is plotted as a function of the average number of allocators contacted by recipients, with separate curves showing its behavior for various the fraction of resource allocators.
Overall, the trends in the curves of Fig.~\ref{fig6}(b) and Fig.~\ref{fig6}(d) resemble those of Fig.~\ref{fig2}(b) and Fig.~\ref{fig2}(d), respectively.
When the disease has a high prevalence (with a small $\langle k_2\rangle$), the trends exhibited by the curves in Fig.~\ref{fig6}(a) and Fig.~\ref{fig6}(c) correspond to those in Fig.~\ref{fig2}(a) and Fig.~\ref{fig2}(c), respectively.
We once again observed four different behaviors of $\rho$, namely, {\it monotonically increasing}, {\it monotonically decreasing}, {\it initially decreasing and then increasing}, and {\it a sudden decrease with large fluctuations}.
The bistable behavior is also observed in the parameter region after the fourth pattern.
The primary disparity between Fig.~\ref{fig6} and Fig.~\ref{fig2} is that as $\langle k_2\rangle$ increases, the disease will eventually disappear in all four scenarios. 
This phenomenon can be attributed to the setting of the infection probability.

\subsection{Comparison of the theory and the simulation}
\begin{figure}
\center
	\includegraphics[width=0.5\linewidth]{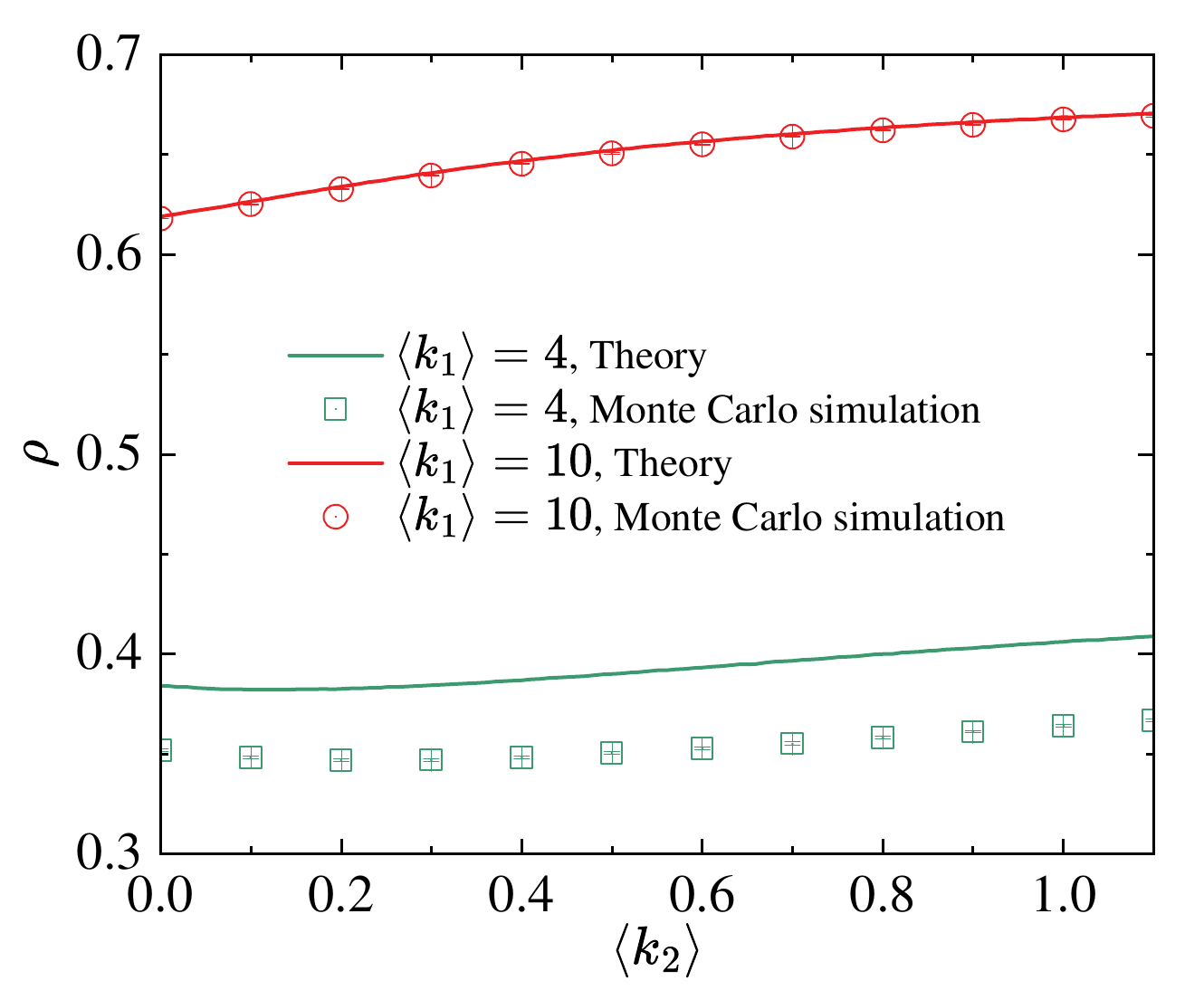}
	\caption{A comparative analysis of the simulation results with the theoretical results. The prevalence $\rho$ as a function of the average number of connections $\langle k_2\rangle$ from recipients to allocators.
The lines are the numerical solutions obtained from Eqs.~\eqref{tD} and \eqref{tA}, while the scatters are the simulation results. Parameters: $\beta=0.12$, $\mu_0=0.5$, $\alpha=1$, $\omega=0.1$, $r=0.1$, and $N=10^5$.}\label{fig7}
\end{figure}
Within the resource layer, we define $k_{\mathrm{max}}^{\mathrm{R\tilde{G}}}$ as the maximum number of \~{D} individuals to which a single \~{G} individual can be connected, and $k_{\mathrm{max}}^{\mathrm{R\tilde{D}}}$ as the maximum number of \~{G} individuals that can be connected to a single \~{D} individual (i.e., the maximum degree of \~{D} individuals).

To analyze the time complexity of Eq.~\eqref{p} and Eq.~\eqref{P}, we deploy a theoretical probe---a node \~{G} configured with $k_{\mathrm{max}}^{\mathrm{R\tilde{G}}}$ links to \~{D} individuals of degree $k_{\mathrm{max}}^{\mathrm{R\tilde{D}}}$.
Upon analysis, it becomes evident that the time complexity of Eq.~\eqref{p} is $O(2^{k_{\mathrm{max}}^{\mathrm{R\tilde{D}}}})$. This exponential complexity poses significant challenges in terms of implementation.
In this context, we can take into account the following polynomial multiplication
\begin{equation}\label{p2}
  \prod_{i \in \Omega (j')\backslash\{j\}}(\rho _i^{ \mathrm{\tilde{G}S} }+\rho _i^{ \mathrm{\tilde{G}I} }x)=\sum_{k}p\left(\frac{1}{r(k+1)}\right)x^k,
\end{equation}
whose coefficients exactly represent the probability distribution as defined by Eq.~\eqref{p}.
At this stage, the time complexity has been reduced from $O(2^{k_{\mathrm{max}}^{\mathrm{R\tilde{D}}}})$ to $O\left((k_{\mathrm{max}}^{\mathrm{R\tilde{D}}})^2\right)$.
For the convolution of Eq.~\eqref{P}, the time complexity is $O \left( \left( k_{\mathrm{max}}^{\mathrm{R\tilde{D}}}\right)^{k_{\mathrm{max}}^{\mathrm{R\tilde{G}}}}\right)$. This time complexity is related to $k_{\mathrm{max}}^{\mathrm{R\tilde{D}}}$ and $k_{\mathrm{max}}^{\mathrm{R\tilde{G}}}$, especially the exponential growth of $k_{\mathrm{max}}^{\mathrm{R\tilde{G}}}$.

Moreover, both Eq.~\eqref{p} and Eq.~\eqref{p2} involve the successive multiplication of a substantial quantity of numbers between 0 and 1, with the number of multiplications being directly proportional to $k_{\mathrm{max}}^{\mathrm{R\tilde{D}}}$.
When a program performs a substantial number of cumulative multiplications with values in the interval $(0,1)$, the following issues may arise: numerical overflow exceptions, precision loss, and rounding errors, among other related concerns.
Specifically, the product may fall below the minimum representable value, triggering an underflow and resulting in either zero or an anomalous output.

In light of the foregoing analysis, we are unable to offer theoretical comparison values for all the simulation data presented in Sec.~\ref{reb} and Sec.~\ref{rec}.
For example, in the network corresponding to $r=0.01$ and $\langle k_2\rangle=10$ in Fig.~\ref{fig2}, it is verified through checking the original data that $k_{\mathrm{max}}^{\mathrm{R\tilde{G}}}$ and $k_{\mathrm{max}}^{\mathrm{R\tilde{D}}}$ are $26$ and $1078$ respectively.
In this subsection, we conduct numerical calculations on the networks where $r=0.1$ and $\langle k_2\rangle\in[0,1.1]$, see Fig.~\ref{fig7}.

As shown in the results of $\langle k_1\rangle=4$ in Fig.~\ref{fig7}, the curve trend presented by the theoretical results is consistent with the simulation results, with the absolute error $\epsilon$ less than $0.042$. 
Here, $\epsilon$ represents the absolute difference between the theoretical and the simulation values.
In a static network, the states of nodes are dynamically correlated~\cite{PhysRevLett.116.258301,PhysRevE.103.032313}. However, the microscopic Markov chain theory fails to consider these dynamic correlations~\cite{RevModPhys.87.925}, which accounts for the existing gap.
While several theoretical methods, such as the epidemic link equations approach~\cite{matamalas2018effective} and the effective degree Markov-chain approach~\cite{PhysRevResearch.5.013196}, do capture dynamical correlations, the resultant systems exhibit a substantial increase in the number of equations.

As the average degree of the network increases, it is well established that the dynamical correlation decreases~\cite{PhysRevResearch.6.L022017}.
In Fig.~\ref{fig7}, when the value of $\langle k_1\rangle$ is raised to $10$, one can observe a remarkable decline in the absolute error, with $\epsilon$ being less than 0.002.
Meanwhile, it is evident that as $n$ increases, the theoretical results progressively align with the simulation results.

\section{Conclusion}\label{conclusion}
In summary, we propose a resource-epidemic dynamics model based on role division and heterogeneous resource allocation.
Under the role division, network nodes are categorized into two classes: resource recipients and resource allocators.
The heterogeneity in resource allocation arises from resource allocators adopting one of two distinct strategies: standard resource distribution or no resource distribution.
Specifically, resource allocators cease resource distribute upon becoming infected.
This mechanism mirrors real-world scenarios, such as the temporary closure of small clinics or the implementation of lockdown measures in hospitals during an epidemic.

In this paper, we investigate how two key parameters affect epidemic spread: (i) the fraction $r$ of resource allocators, and (ii) the average number $\langle k_1\rangle$ of resource allocators directly connected to resource recipients.
Meanwhile, based on the baseline recovery probability and the treatment efficiency, we establish the relation between resources and recovery probability. 
In our investigative analysis, four combinations are examined, where each parameter ($\alpha$, $\omega$) is set to low or high levels.
We find four distinct behaviors of prevalence $\rho$ as the parameter $\langle k_1\rangle$ increases: monotonically increasing, monotonically decreasing, initially decreasing and then increasing, and a sudden decrease with large fluctuations.
Meanwhile, the bistable region emerges after the phase marked by a sudden decrease with large fluctuations.

Among these phenomena, one characterized by a pattern of initially decreasing and then increasing is the most prevalent, observed in all four configuration scenarios. 
The fundamental reason behind this phenomenon is that as $\langle k_2\rangle$ increases, the total resources available to  infected recipients (\~{G}I individuals) initially rise and then fall. 
Fundamentally, this embodies a competitive dynamic wherein an increase in the number of links (edges) between recipients and allocators not only expands the channels for resource allocation but also increases the infection risk of the allocators.

As mentioned in the above competitive relation, there exists a certain window for the parameter $\langle k_2\rangle$ during which the number of infected recipients decreases, while that of infected allocators increases. 
Moreover, as $r$ increases, it enhances the proportion of infected allocators in the overall population.
On the other hand, when parameters  $\alpha$ and $\omega$ remain at low levels, they weaken the reduction in infected recipients. 
These combined factors collectively result in the emergence of the monotonically increasing curve.

At high levels of medical resource efficiency, only minimal resources are required to achieve saturation in recovery probability. 
Consequently, significant redundant resources accumulate within the population. 
Under these conditions, augmented inter-node connectivity (between recipients and allocators) facilitates uniform resource distribution, thereby enhancing overall resource use efficiency.
These effects explain the emergence of monotonically decreasing.

Finally, we observe that the presence of a small number of \~{G}I individuals can initiate a cascade, suggesting their critical role in driving large-scale transitions.
This mechanism explains the emergence of a sudden decrease with large fluctuations, and we further identify a bistable behavior in the system, where the final state depends on the initial seeds.

In theory, we extend the microscopic Markov chain to our model.
Notably, by means of convolution, we precisely calculate the average recovery probability. 
Our findings show that when the average links among resource recipients is high, the theoretical results are in excellent agreement with the simulation results.

In this paper, some of the assumptions effectively isolated the key parts of the resource allocation process for study, which also provides an opportunity to discuss more complex real-world scenarios in future research. 
On the one hand, in the modern healthcare system, the allocators can achieve resource sharing through means such as referrals (meaning that a small number of \~{D}-\~{D} links may exist), which will also make the resource distribution more even.
On the other hand, the operation mode of clinics does indeed satisfy the assumption that resource allocation is completely halted immediately after the allocator is infected. However, hospitals can operate even with reduced capacity. Therefore, a gradual failure mechanism is also worth exploring.
Finally, in line with the majority of previous research, our role-based resource allocation approach did not incorporate the cost of resources. 
Once resource costs were taken into account, not all individuals would engage in treatment proactively and cooperatively. 
Therefore, it is essential to conduct research on the co-evolution of cooperation and disease transmission~\cite{PhysRevResearch.7.023211}.
Meanwhile, individuals will form highly individualized and distinct perceptions of the same resource cost based on their own biases and experiences~\cite{tgkt-54gc}.

\appendix 
\renewcommand{\thesection}{Appendix}
\section{Detailed explanation of the selection of parameters $\alpha$, $\omega$, and $\beta$}\label{supa}
\begin{figure}
\includegraphics[width=\linewidth]{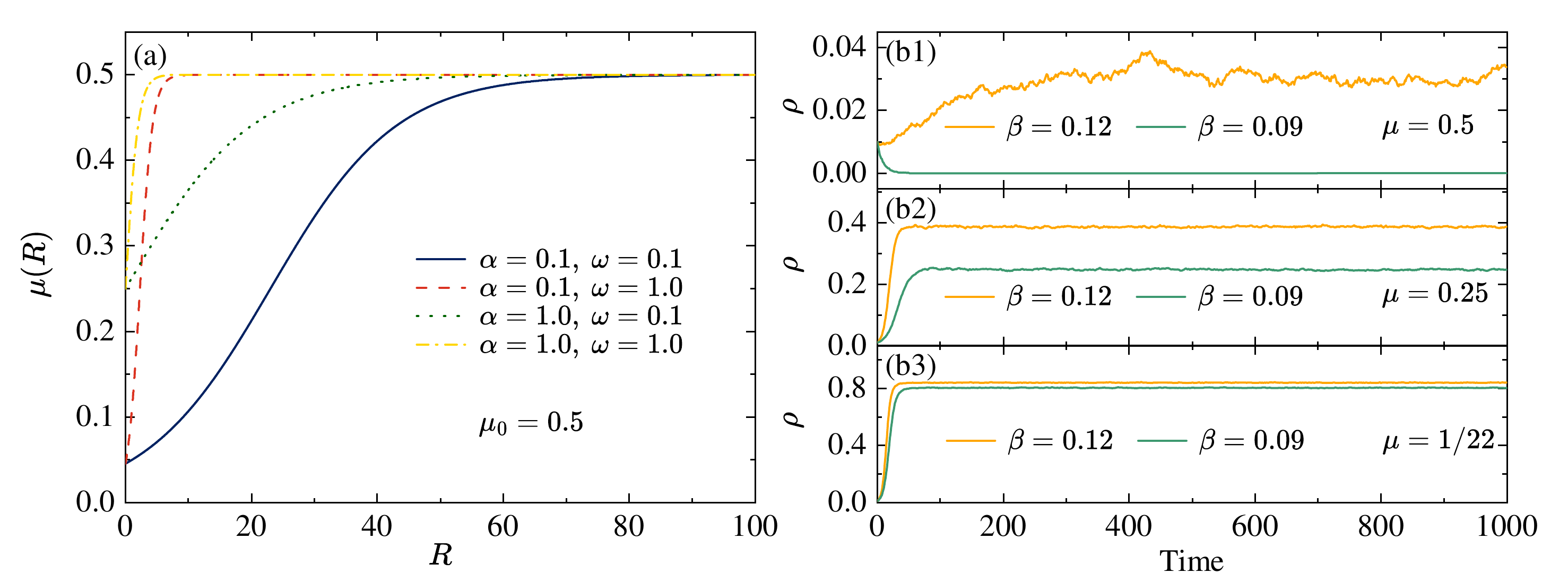}
\caption{(Color online) (a) Solutions of the Eq.~\eqref{mu}. 
(b1)-(b3) Time series for the prevalence. The network is the same as the one in Fig.~\ref{fig2} with parameters $r=0.01$ and $\langle k_2\rangle=0$.
\label{figapp}}
\end{figure}

For the baseline recovery probability and the treatment efficiency, we have considered both weak and strong scenarios. Specifically, we have investigated four combinations: weak baseline probability coupled with weak  treatment efficiency ($\alpha=0.1$ and $\omega=0.1$), weak baseline probability combined with strong medical  treatment efficiency ($\alpha=0.1$ and $\omega=1.0$), strong baseline probability paired with weak  treatment efficiency ($\alpha=1.0$ and $\omega=0.1$), and strong baseline probability accompanied by strong  treatment efficiency ($\alpha=1.0$ and $\omega=1.0$).
Based on the relation between the recovery probability and $\alpha$, $\omega$ in Eq.~\eqref{mu}, Fig.~\ref{figapp}(a) shows the solutions corresponding to these combinations.
As $R$ increases, $\mu$ gradually approaches the saturation value [$\mu(R\to\infty)=0.5$] for $\omega=0.1$, whereas it rises sharply for $\omega=1.0$.
The parameter $\alpha$ governs the recovery probability in the absence of resources. Specifically, when $\alpha=0.1$, the recovery probability $\mu(R=0)$ is $1/22$, whereas for $\alpha=1.0$, $\mu(R=0)$ is $0.25$.

For a homogeneous network, the largest eigenvalue of its adjacency matrix is close to the average degree of the network~\cite{PhysRevE.86.041125}.
Combining Eqs.~\eqref{betac0}-\eqref{betac1}, we we tested and ultimately selected the parameter values $\beta=0.12$ and $\beta=0.09$ for this study.
Figure~\ref{figapp}(b1)-(b3) presents the time series of the prevalence $\rho$ under uniform recovery probability for all individuals. 
Specifically, Fig.~\ref{figapp}(b1) represents the scenario where $R\to\infty$, Fig.~\ref{figapp}(b2) corresponds to $R=0$ with $\alpha=1.0$, and Fig.~\ref{figapp}(b3) pertains to $R=0$ with $\alpha=0.1$.
It is evident that when $\beta=0.12$, the disease will break out even in the presence of abundant resources. 
However, when $\beta=0.09$, the disease fails to spread under conditions of abundant resources, yet it will spread when there are no resources.

\printcredits

\section*{Declaration of Competing Interest}
The authors declare that they have no known competing financial interests or personal relationships that could have appeared to influence the work reported in this paper.

\section*{Acknowledgments}
This work was supported by the Natural Science Basic Research Plan in Shaanxi Province of China (Grant No.\ 2025JC-YBMS-019) and the National Natural Science Foundation of China (Grant No. 12231012).

\section*{Data availability}
All data supporting the findings of this study are available within the paper, and our code is available for public access at https://github.com/chaoran-cai/CCR-JHX-25.

\bibliographystyle{model1-num-names}

\bibliography{ref}

\end{document}